# Accurate Nanoscale Mapping of Electric Fields across Random Grain Boundaries in Polycrystalline Oxides Using Precession-Assisted 4D-STEM

Sangjun Kang[1, 2, 3*], Hyeyoung Cho[1, 2], Maximilian Töllner[1, 2], Anna Rose Nelson[1], Ziming Ding[2], Xiaoke Mu[4], Di Wang[2, 3], Wolfgang Rheinheimer[5], Kai Wang[6], Bai-Xiang Xu[6], Jakob Konstantin Laux[7], Mahmoud Serour[7], Karsten Albe[7], Andreas Klein[8], Christian Kübel[1, 2, 3*]

[1]*In situ Electron Microscopy, Department of Materials Science, Technical University of Darmstadt (TUDa), 64287 Darmstadt, Germany*
[2]*Institute of Nanotechnology (INT), Karlsruhe Institute of Technology (KIT), 76344 Eggenstein-Leopoldshafen, Germany*
[3]*Karlsruhe Nano Micro Facility (KNMFi), Karlsruhe Institute of Technology (KIT), 76344 Eggenstein-Leopoldshafen, Germany*
[4]*School of Materials and Energy and Electron Microscopy Centre, Lanzhou University, Lanzhou 730000, China*
[5]*Institute for Ceramic Materials and Technologies, University of Stuttgart, Allmandring 7B, 70569 Stuttgart, Germany*
[6]*Mechanics of Functional Materials Division, Institute of Materials Science, Technische Universität Darmstadt, Darmstadt, 64287, Germany*
[7]*Materials Modeling, Department of Materials Science, Technical University of Darmstadt (TUDa), 64287 Darmstadt, Germany*
[8]*Electronic Structure of Materials, Department of Materials Science, Technical University of Darmstadt (TUDa), 64287 Darmstadt, Germany*

*Corresponding authors: sangjun.kang@tu-darmstadt.de and christian.kuebel@kit.edu*

**Abstract:** Space charge layers (SCLs) at grain boundaries play a crucial role in modulating local electric fields and influencing the functional properties of materials, such as oxygen vacancy migration and ionic conductivity in oxide ceramics. However, the direct experimental analysis of such localized electric fields and the corresponding charge distribution remains challenging. Conventional center-of-mass (CoM) analysis in scanning transmission electron microscopy differential phase contrast (STEM-DPC) is strongly affected by orientation-dependent contrast and dynamical scattering. Here, we demonstrate that combining electron beam precession with advanced post-processing, employing iterative edge detection via a Sobel filter and singular value decomposition (SVD), enables reliable and accurate, unbiased diffraction shift measurements with minimal crystallographic artefacts. The new method accurately refines the central disk position in nanobeam electron diffraction (NBED) patterns and thus significantly improves the extraction of the local electric field and corresponding charge distribution. Comparative analysis with conventional CoM methods shows superior accuracy and robustness for random grain boundaries in BaTiO$_3$ and SrTiO$_3$ as exemplary case studies. The experimental work is complemented by atomistic simulations to separate the electric field of the SCL from the mean inner potential difference of the grain boundary and the elemental segregation around the grain boundary. The in-depth analysis shows that our approach enables high-fidelity mapping of electromagnetic fields and their charge distribution in complex polycrystalline specimens, laying the groundwork for improved quantitative analysis using STEM-DPC.

**Introduction**

The SCL around grain boundaries in non-conducting ceramics critically modulates the local electric field and thereby influencing key functional properties of materials, such as oxygen vacancy migration and ionic conductivity in oxide ceramics. [1-8] Quantitative mapping of the local electromagnetic field is essential not only for understanding fundamental mechanisms, but also for advancing applications in electronic devices, magnetic storage media, and quantum materials. [3, 9-11] Various techniques have been explored for high-resolution field mapping, including off-axis electron holography and STEM-DPC imaging. [12-20] Off-axis electron holography can directly record the phase shift of the electron wave to visualize electrostatic potentials and magnetic induction, but it requires a specialized TEM with a biprism and a coherent reference wave within the field of view, [12-14] and in practice suffers from phase-wrapping artifacts and strict noise/dose constraints. As a complementary approach, STEM-DPC measures the beam deflection by determining either the rigid shift of the diffraction pattern or the redistribution of intensity within the direct beam, offering real-space field maps without the need for a coherent reference wave. [21-25] However, traditional STEM-DPC implementations using a segmented detector have limited sensitivity and are prone to background drifts. Even more critical, diffraction contrast from crystallographic features imposes significant systematic, orientation-dependent shifts, making it challenging to identify the true electric field signal [26]. These challenges become particularly pronounced in heterogeneous specimens, where asymmetric or randomly oriented grain boundaries, local thickness variations, and crystallographic misorientations lead to inhomogeneous bright-field disk intensities that distort the true DPC signal.

Four-dimensional scanning transmission electron microscopy (4D-STEM) equipped with high-angular-resolution pixelated detectors has enabled improved data acquisition and processing for visualizing internal electric fields in materials. [15, 26-28] By recording a full diffraction pattern at every probe position, 4D-STEM can capture subtle beam deflection caused by local electric and magnetic fields. [29-31] However, accurately extracting the electric field components from 4D-STEM data remains challenging, as diffraction contrast variations arising from orientation, thickness, and multiple scattering often obscure the field-induced shifts. To address these limitations, Nakamura et al. proposed a tilt-series DPC method, where multiple

STEM-DPC images are acquired at slightly varied specimen tilts. [32] By averaging these images, orientation-sensitive diffraction contrast can be reduced while reinforcing the tilt-insensitive electromagnetic field signals. This approach has been successfully applied to semiconductor p–n junctions and ceramic bicrystals. [15, 30] Alternatively, electron precession can be employed to reduced diffraction contrast and multiple scattering. [18, 24, 33] However, even high precession angles of ~1° are not sufficient to fully eliminate diffraction contrast in polycrystalline ceramics, which typically exhibit broad orientation distributions and complex grain boundary networks. In this case, CoM fitting, widely used to track disk shifts in 4D-STEM, can suffer from serious artefacts in polycrystalline materials due to remaining asymmetric intensity distributions within the central diffraction disk, caused by local misorientations and dynamic scattering. [33] To overcome this, recent studies introduced a 4D-STEM approach that uses edge detection algorithms to extract the disk boundary more reliably from diffraction patterns. [28, 31, 33-35] While these methods reduce intensity redistribution effects caused by the local crystal orientation and thickness and thus allow more accurate disk tracking, they were only demonstrated for systems with strong DPC signals, such as semiconducting junction, micrometer-scale Landau domains or vortex structures in magnetic materials. To date, a systematic investigation of the DPC signal accuracy, artifact suppression, and thus the reliability to measure small diffraction displacements in polycrystalline materials has not been reported. Especially for the subtle nanometer scale electric field variations in SCLs around random grain boundaries, this accuracy of the measurement and artifacts suppression is absolutely critical.

In this study, we overcome previous limitations and extend electric field mapping across random grain boundaries in polycrystalline oxides. We integrate electron beam precession with a robust post-processing workflow involving Sobel edge detection of diffraction disk edges and SVD-based disk center refinement. This combined approach enables reliable separation of genuine field-induced beam deflections from diffraction contrast artifacts and achieves sub-pixel precision in measuring disk shifts. We demonstrate that our method can extract high-precision maps of electric fields in materials with complex nanoscale structure, including polycrystalline samples that were previously intractable. The result is a significant improvement over conventional CoM-based DPC, broadening the scope of 4D-STEM for quantitative electrostatic and magnetic field analysis in heterogeneous systems. The refined field measurements can be used to derive the electrostatic potential and the charge density distribution as related physical properties. The interpretation of the experimental measurements is supported by complementary atomistic simulations showing that the measured response is dominated by the electric field distribution of the space-charge-layer, providing deeper insights into SCLs in polycrystalline oxides. Our results suggest that a full quantitative interpretation of the measurements requires explicit consideration of the segregation-induced contribution to the mean inner potential within the space-charge layer (SCL).

# Methods

## 2.1 Sample preparation

Fe-doped BaTiO$_3$ (BTO) bulk ceramics were prepared by conventional solid-state reaction using stoichiometric mixtures of BaCO$_3$ and TiO$_2$ with Fe$_2$O$_3$ as the Fe precursor. The nominal Fe content was 2.4 at% on the Ti site. The precursor powders were ball milled in ethanol for 24 h. After drying, the powder mixture was calcined in air at 1250 °C for 2 h. The calcined powder was uniaxially pressed into pellets (10 mm in diameter and approximately 1.5 mm in thickness) at 200 MPa and sintered in air at 1450 °C for 4 h. Fe-doped SrTiO$_3$ (STO) bulk ceramics were prepared by conventional solid-state reaction using stoichiometric mixtures of SrCO$_3$ and TiO$_2$ with Fe$_2$O$_3$ as the Fe precursor. The nominal Fe content was 2 at% on the Ti site. The precursor powders were ball milled in isopropanol with 3 mm ZrO$_2$ balls using a PM400 planetary ball mill for 15 cycles of 5 min at 400 rpm. The powders were sieved and calcined at 975 °C for 3 h, followed by a second milling step of 6 cycles of 5 min at 300 rpm. Green bodies were uniaxially pre-pressed, cold-isostatically pressed at 400 MPa, and sintered in air at 1350 °C for 1 h. The furnace heating and cooling rates were 20 and 10 K min$^{-1}$.

Thin lamellae for TEM analysis were prepared using focused ion beam (FIB) milling using a Strata 400 S (FEI) from the bulk oxides. Final thinning was performed to achieve ~-50-100 nm thickness for electron transparency at the area of interest, employing a stepwise reduction in accelerating voltage from 30 kV to 5 kV and beam currents from 8 nA to 2 pA to minimize Ga$^+$ ion-induced damage. Selected lamellae were further polished using a NanoMill (Fischione) to 500 eV.

Large-area STO samples were prepared by conventional TEM specimen preparation. The sintered ceramics were cut and dimpled, followed by ion polishing using a PIPS II ion mill (Gatan) with a final milling energy of 200 eV.

## 2.2 Atomic Structure Simulations

Atomic configurations of a BaTiO$_3$ Σ5 tilt boundary were constructed using MD simulations implemented in the LAMMPS package. The initial model consisted of a bicrystal supercell with a Σ5(310)[001] symmetric tilt boundary placed at the center of a 25 nm cubic simulation box. The bicrystal was generated by joining two single-crystal BTO slabs with a relative rigid-body translation to identify the lowest-energy grain boundary configuration. Interatomic interactions were described using atomic cluster expansion (ACE) machine learning potential [36] parameterized for BaTiO$_3$ using the PACEMAKER code [37] to accurately reproduce the lattice dynamics, ferroelectric properties, and defect energetics of the material. After initial construction, the system was relaxed using energy minimization and subsequently equilibrated under the NPT ensemble at 700 K and 0 bar using the Nosé–Hoover thermostat and barostat to ensure thermal and mechanical stability of the structure. To obtain a well-relaxed grain boundary configuration, multiple grain boundary translations along the boundary plane were tested, and the configuration with the lowest potential energy was selected. Structural relaxation was followed by additional short MD runs at room temperature to allow for local reconstruction and to minimize residual stresses near the boundary. The final atomic structure was analyzed to investigate coordination changes and defect energies near the grain boundary.

## 2.3 Experiment-driven atomistic model modification, mean inner potential and space charge layer

The atomic structure described above was used as the initial atomic model. The simulation cell (785.20955 × 12.62398 × 3.99205 Å$^3$) was replicated to implement compositional modification resulting in a supercell of 785.20955 × 782.68697 × 99.80135 Å$^3$. Two atomic models were constructed for subsequent simulations, an undoped reference model and a Fe-doped model. To construct the Fe-doped model, Fe segregation and oxygen defect concentrations were introduced to resemble the experimental STEM-EDS composition in the SCL (Figure S1), with the observed segregation width of 10 nm to define a Gaussian spatial weighting function. Fe segregation was implemented by substituting Fe atoms for Ti according to this weighting function, reproducing the measured Fe/Ti ratio of 0.064 next to the grain boundary and 0.021 in the bulk. The oxygen deficiency at the grain boundary is present from the MD model. Additional oxygen deficiency in the bulk was introduced by probabilistic removal of O atoms, with a higher removal probability in the bulk than in the SCL, to mimic the oxygen-vacancy depletion expected in a conventional positive-core space-charge layer. The target oxygen distribution was guided by the defect-chemistry-informed phase-field model of Wang et al. [38].

The mean inner potential for the two structures has been calculated in abTEM [39] with the independent atom model. The projected potential has been convoluted with a Gaussian profile with a diameter of the lattice parameters to reflect to experimental electron beam diameter and further averaged along the grain boundary to calculate the mean inner potential profiles across the boundary. Analogously, line profiles for atomic fractions for the different elements have been calculated by projection the atomic counts along the grain boundary and convolution with the Gaussian profile followed by normalization with respect to the bulk atomic density.

The space-charge potential used in the present analysis was adopted from the continuum restricted-equilibrium model reported by Usler et al. [40] for acceptor-doped SrTiO$_3$. In that work, the space-charge potential was obtained as the grain-boundary-to-bulk potential difference by solving the Poisson equation self-consistently with defect-chemical relations and site-exclusion constraints, while accounting for oxygen-vacancy segregation at the grain-boundary core and frozen-in acceptor and oxygen-vacancy distributions under restricted-equilibrium conditions.

## 2.4 STEM and 4D-STEM experiment

Conventional STEM measurements were performed using a probe-corrected Themis 300 (TFS) at an operation voltage of 300 kV in microprobe STEM mode. Spot size 5 and a semi-convergence angle of 2.5 mrad were employed, resulting in a diffraction-limited probe size of approximately 0.5 nm. HAADF images were acquired in this mode with a camera length of 380 mm and a collection angle of 21–125 mrad.

EDS measurements were conducted using a Super-X G2 detector system integrated within the Themis 300 microscope. The same beam settings were employed for HAADF imaging. Spectral data were acquired with a dwell time of 40 μs per pixel. The detector was operated with a dispersion of 5 eV, an energy offset of −250 eV, and a shaping time of 3 μs. All four segments of the Super-X detector were active during acquisition, ensuring high solid angle coverage.

For the 4D-STEM measurements, different experimental conditions were applied depending on the measurement purpose. For electric field measurement, a convergence angle of 2.5 mrad was employed to prevent overlap between the diffraction disks and the direct beam, ensuring clear separation for accurate analysis. Diffraction patterns were recorded using a Quadro detector (DECTRIS) at a camera length of 3.1 m to magnify the direct beam and enhance DPC sensitivity. The acquisition time was set to 1 ms per probe position, with a scan step size of 5 Å for higher resolution imaging focused on the grain boundary and 5 nm for lower resolution large field of view mapping. One 4D-STEM data set was acquired without precession and a second corresponding one with precession applied using a Nanomegas system with a precession angle of 0.6°, performed with a single rotation per probe position. The orientation map was processed using the commercial ASTAR software (Nanomegas). For measuring the strain field, a smaller convergence angle of 0.5 mrad and a camera length of 460 mm were employed to achieve higher angular resolution in the diffraction patterns. A step size of 5 Å across the grain boundary was used in this configuration.

## 2.5 Data processing for DPC mapping

To retrieve the projected electric field from 4D-STEM datasets, we analyzed the shift of the central bright-field (BF) disk throughout the scan using a geometry-based edge fitting method. This approach enables precise determination of beam deflection by estimating displacement of the disk with sub-pixel accuracy.

Each diffraction pattern was preprocessed by applying a two-dimensional Hanning window, which was logarithmically scaled and normalized to enhance the visibility of the central disk while suppressing background and boundary artifacts. An optimized Sobel edge detection filter was applied to identify the boundary of the bright-field disk. A pattern-specific adaptive thresholding algorithm was employed, where the threshold value was dynamically optimized for each diffraction pattern to yield a clean and continuous edge contour without including spurious features. Unlike methods that rely on the intensity distribution or diffraction ring shape, this edge-based method directly captures the geometric boundary of the central disk [31, 34]. This strategy reduces the influence of diffraction contrast and crystal orientation effects, allowing for more accurate measurement of rigid beam shifts. This is particularly advantageous in regions where intensity asymmetries or diffraction contrast can bias CoM-based measurements.

The detected edge coordinates were fitted to an ellipse using an algebraic conic fitting algorithm based on SVD. The ellipse was modeled using the general second-order form:

$$\mathbf{F(C, Q)} = \mathbf{C} \cdot \mathbf{Q} = aq_x^2 + bq_xq_y + cq_y^2 + dq_x + eq_y + f,$$

where $\mathbf{C} = [a\ b\ c\ d\ e\ f]$ and $\mathbf{Q} = [q_x^2\ q_xq_y\ q_y^2\ q_x\ q_y\ 1]^T$. $\mathbf{F(C, Q)}$ is the mismatch distance of a data point $(q_x, q_y)$ to the ellipse $\mathbf{F(C, Q)} = 0$. Thus, the best fit of an ellipse to the edge of the diffraction disk is equivalent to finding C to minimize the sum of squared mismatch distances $\mathbf{D(C)} = \min \sum_{i=1}^{N} \mathbf{F}(C_i, Q_i)^2$, where $N$ is the total number of selected pixels by the thresholding, and $i$ is the pixel sequence number. This least square problem can be solved by SVD considering a rank-deficient generalized eigenvalue system, as $\mathbf{QQ^TC^T} = \lambda \mathbf{PC^T}$, where $\mathbf{P}$ is a constrain matrix to avoid trivial solutions, e. g. $\mathbf{C} = 0$, as defined by

$$\mathbf{P} = \begin{bmatrix} 0 & 0 & -2 & 0 & 0 & 0 \\ 0 & 1 & 0 & 0 & 0 & 0 \\ -2 & 0 & 0 & 0 & 0 & 0 \\ 0 & 0 & 0 & 0 & 0 & 0 \\ 0 & 0 & 0 & 0 & 0 & 0 \\ 0 & 0 & 0 & 0 & 0 & 0 \end{bmatrix}.$$

The fitted parameters were converted into geometric quantities (center position and radius). The displacement of the fitted ellipse center relative to a reference position was used to compute the beam deflection vector at each scan point, which was linearly mapped to the projected in-plane electric field. To compensate for scan pivot point errors, a reference dataset was acquired from a vacuum region using the same optical settings prior to the actual measurements, and the resulting systematic offset was subtracted from the sample data.

## 2.6 Cepstral analysis for strain mapping

Strain mapping was carried out using cepstral analysis applied to each diffraction pattern in the 4D-STEM dataset. [41] This method utilizes the Fourier transform of diffraction features in reciprocal space to extract local lattice parameter variations with high spatial resolution. Specifically, the logarithm of the Fourier transform of each diffraction pattern was computed to obtain the cepstrum, which emphasizes periodic modulations corresponding to the underlying lattice spacings. A band-pass filter was applied in the cepstral domain to isolate the frequency components associated with the Bragg disk spacing. The resulting filtered signal was then analyzed to extract the positions of the cepstral peaks, which reflect the periodicity and orientation of the diffraction pattern. By measuring shifts and distortions in these cepstral features across the scan, local strain components were derived. Compared to direct-space geometric methods, e.g. template matching or Hough transformation, cepstral strain mapping provides enhanced robustness against variations in intensity, crystal orientation, and probe aberrations. Since the analysis is based on the periodicity of reciprocal lattice information rather than the absolute position or shape of individual Bragg disks, it is less sensitive to local orientation variation and diffraction contrast. Moreover, it significantly reduces errors associated with excitation conditions, as it does not rely on satisfying the exact Bragg condition at each scan point. This allows for more accurate and stable strain quantification even in regions with strong orientation gradients, grain boundaries, or slight misalignments. The analysis was performed using the MATLAB-based software package developed by Padgett et al. [41], which implements the full cepstral workflow including Fourier transformation, filtering, peak detection, and strain tensor reconstruction.

To further correlate the local strain state with variations in the mean inner potential (MIP), we computed the volumetric strain $\varepsilon_{Vol}$ from the reciprocal lattice vector deformation assuming the third dimension to be uniform. As illustrated in Figure 1e, two non-collinear reciprocal lattice vectors $g_1$ and $g_2$ were selected from each diffraction pattern, and their local strain components were calculated as $\varepsilon_{g1} = \frac{|g_1 - g_{1,0}|}{g_{1,0}}$ and $\varepsilon_{g2} = \frac{|g_2 - g_{2,0}|}{g_{2,0}}$, where $g_{1,0}$ and $g_{2,0}$ represent the corresponding vectors in an unstrained reference region. The volumetric strain was then defined as $\varepsilon_{Vol} = \frac{\varepsilon_{g1} + \varepsilon_{g2}}{2}$. This scalar measure reflects the projected change in unit cell area, which is directly related to local atomic density variations. Because the mean inner potential is sensitive to electron density, $\varepsilon_{Vol}$ provides a reliable strain-sensitive descriptor for interpreting MIP contrast observed in DPC. In particular, $\varepsilon_{Vol}$ provides a physically grounded basis for referencing the built-in potential across the sample, allowing for more reliable interpretation of electrostatic potential variations induced by strain or interface effects.

## Results and Discussion

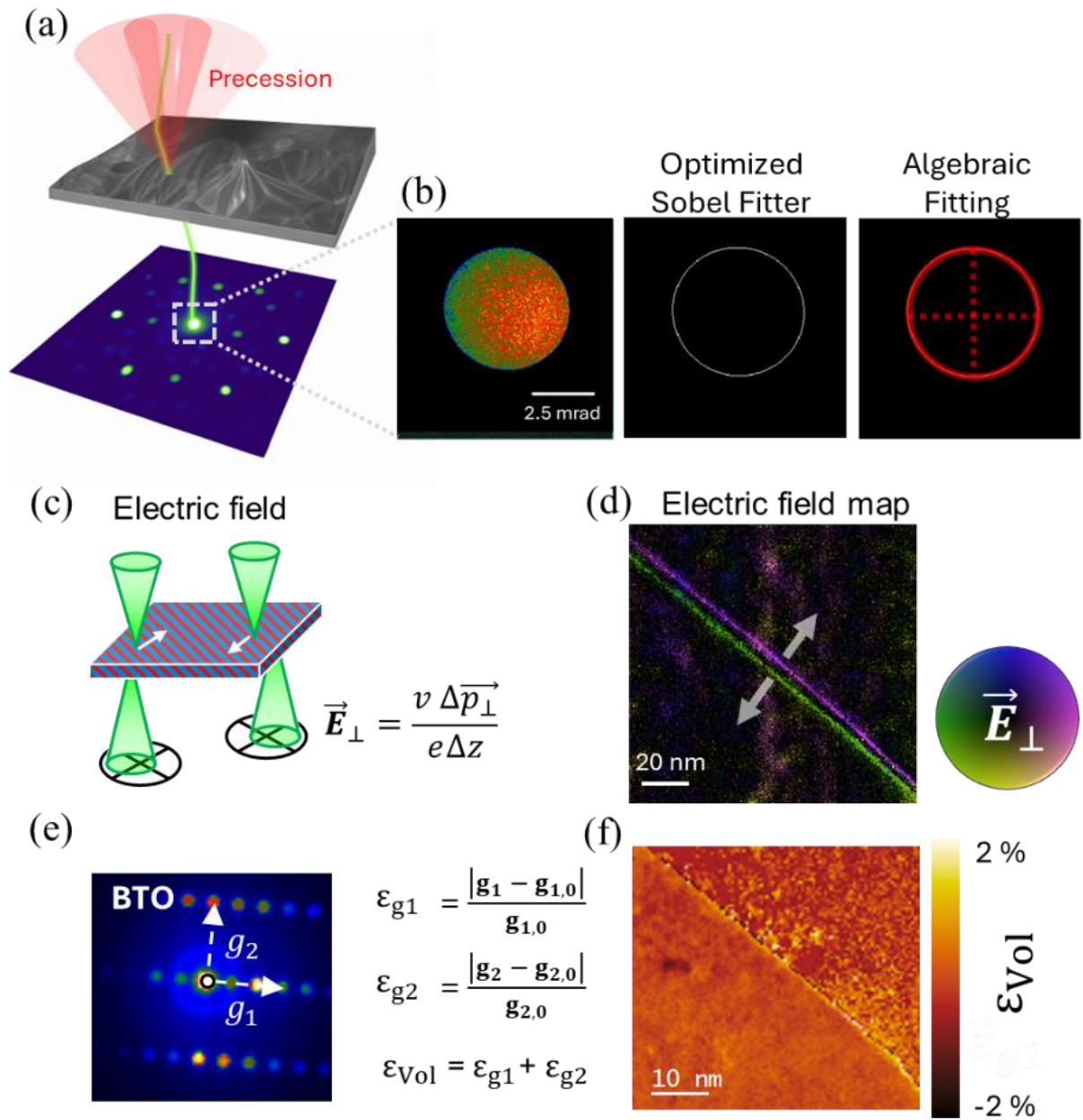

**Figure 1. Workflow for reliable mapping of electric field and strain across grain boundaries in polycrystalline oxides using precession-assisted 4D-STEM and SVD-based diffraction shift detection**. (a) Schematic illustration of electron beam precession during 4D-STEM acquisition to reduce diffraction and orientation contrast. (b) Central bright-field disk from a diffraction pattern processed via an optimized Sobel edge filter followed by algebraic ellipse fitting using SVD to robustly extract the disk position. (c) Electric field-induced deflection of the central disk due to transverse momentum transfer, used to reconstruct the projected electric field. (d) In-plane electric field map showing sharp contrast across a grain boundary; color hue and intensity encode field orientation and magnitude. (e) Definition of reciprocal space strain components from two non-collinear diffraction vectors used to compute volumetric strain. (f) Volumetric strain map across the boundary, serving as a reference to quantify the in-plane MIP gradient due to strain effects at the GB.

Figure 1 outlines the complete workflow used to the map the electric field and strain distribution across a random GB in polycrystalline oxides with high spatial resolution. To suppress strong diffraction contrast and enhance the sensitivity to weak electrostatic signals, we used a precession-assisted 4D-STEM technique (Figure S2). In this approach, the incident electron beam is slightly tilted around a defined cone angle during scanning (Figure 1a), allowing for an angular integration of diffraction patterns that averages out most orientation-related contrast while preserving deflection signals arising from the electric field. The central bright-field (BF) disk in each diffraction pattern was first preprocessed using a 2D Hanning window, followed by logarithmic scaling and normalization to enhance the central disk and suppress background artifacts. An iterative Sobel edge filter with adaptive thresholding was then applied to extract clean and continuous disk boundaries. This edge-based method, which avoids reliance on intensity profiles or diffraction ring shapes, improves robustness against diffraction contrast and orientation effects. The resulting contour was subsequently fitted with an ellipse using SVD, as shown in Figure 1b.

By analyzing the spatial variation of the disk shift across the scanned area, we reconstructed the projected in-plane electric field map (Figure 1d). The map clearly reveals enhanced electric field contrast at the grain boundary, indicating local variations in electrostatic potential associated with the SCL at the boundary. To isolate the contributions of strain-induced potential gradients from those arising from charge accumulation, we simultaneously extracted local strain tensors by cepstral analysis from the 4D-STEM diffraction data using two independent diffraction vectors (Figure 1e). The resulting volumetric strain map (Figure 1f) serves as a baseline to correct for strain induced MIP gradients, allowing to more accurately quantify the true electric field signal associated with SCLs. This combined approach reduces the contribution of various artifacts, enabling more reliable quantification of electrostatic fields at individual grain boundaries in oxide ceramics.

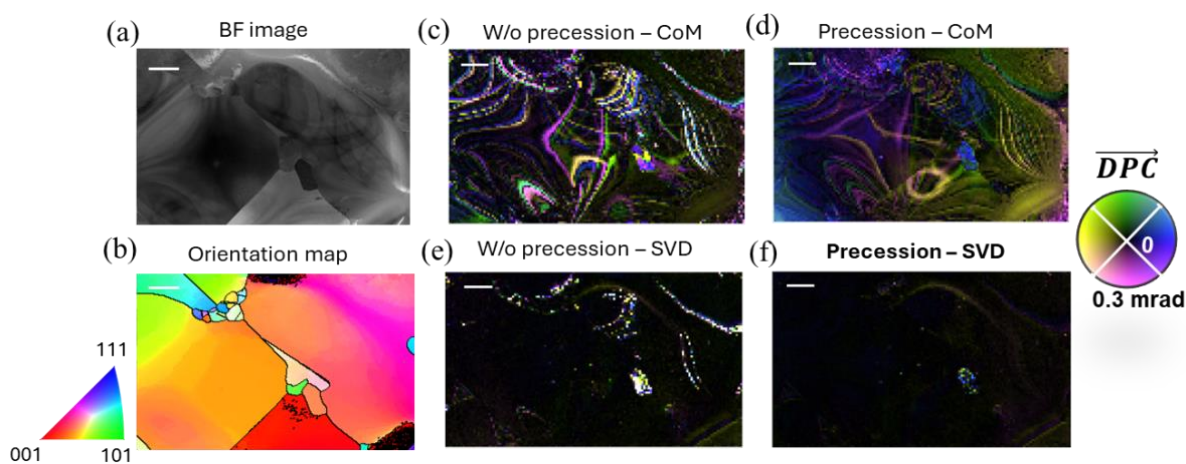

**Figure 2. Comparison of DPC maps obtained by different 4D-STEM acquisition and processing.** (a) BF-STEM image showing the region selected for 4D-STEM acquisition, which is expected to yield a uniform field distribution (except for few GBs). (b) 4D-STEM ACOM orientation map of the same region, showing the local crystallographic orientation distribution. (c, d) Color maps of the DPC signal obtained using the conventional center-of-mass (CoM) method: (c) without beam precession and (d)

with beam precession. (e, f) Color maps of the DPC signal obtained using the SVD-based disk center fitting method: (e) without beam precession and (f) with beam precession. In the DPC maps, hue represents the in-plane direction of the DPC vector and the intensity represents its magnitude. The intensity scale is the same in all DPC maps. All scale bars correspond to 100 nm.

Figure 2 compares DPC maps obtained under different 4D-STEM acquisition and processing conditions in a polycrystalline region. Figure 2a shows the BF-STEM image of the area selected for 4D-STEM acquisition, and Figure 2b shows the corresponding crystal orientation map reconstructed from the 4D-STEM dataset. Except for local variations around the grain boundaries, which are not visible due to the large scan steps used for the overview map, the sample is expected to exhibit a uniform electric field, meaning no field changes should be present. Figures 2c and 2d present the DPC maps obtained using the conventional center-of-mass (CoM) method without and with beam precession. In these maps, hue represents the in-plane direction of the DPC vector and intensity represents its magnitude. The magnitude map is presented in Figure S3.

The CoM map acquired without beam precession exhibits strong spatial contrast associated with diffraction contrast due to local orientation variations, indicating that the measured DPC signal is strongly influenced by an asymmetric intensity redistribution within the BF disk due to diffraction condition variations. When beam precession is applied, the overall background variation is reduced and some of the orientation-dependent contrast is suppressed, showing that angular averaging during acquisition is effective in mitigating part of the diffraction contribution. Nevertheless, the CoM result with precession still exhibit substantial non-uniform contrast, demonstrating that beam precession alone is insufficient to fully remove diffraction-induced artifacts in a polycrystalline specimen. This trend is also reflected quantitatively in the standard deviation measured over the mapped region, which decreases from 0.115 mrad without precession to 0.0166 mrad with precession (histogram in Figure S4). Although this reduction is substantial, the CoM-based result with precession still exhibit a relatively high level of variation, indicating that a significant fraction of the diffraction-induced artifact remains limiting interpretation of the DPC measurements in terms of the local electric field.

Figures 2e and 2f show the DPC maps reconstructed from the same raw data used to generate Figures 2c and 2d using the SVD-based disk fitting method. In contrast to the CoM results, the SVD-based maps are considerably more uniform and show markedly reduced diffraction-related contrast. This improvement is consistent with the geometry-based edge-fitting strategy used for disk shift determination, which is less sensitive to crystal-orientation-dependent intensity redistribution inside the central BF disk than the conventional CoM approach. Even without beam precession, the SVD-based reconstruction substantially lowers the variation of the measured DPC signal, giving a standard deviation of 0.0107 mrad, which is significantly smaller than that of the CoM result under the same condition. When beam precession is combined with SVD fitting, the standard deviation is further reduced to 0.0036 mrad, yielding the most homogeneous background among all cases (histogram in Figure S4). The beam-precessed SVD map in Figure 2f therefore provides the clearest representation of the intrinsic DPC contrast, with minimal artifacts from diffraction-induced background modulation. These results demonstrate that beam precession and SVD-based disk fitting act synergistically: precession suppresses orientation-sensitive diffraction contrast during acquisition, while SVD fitting improves the robustness of disk shift retrieval during post-processing.

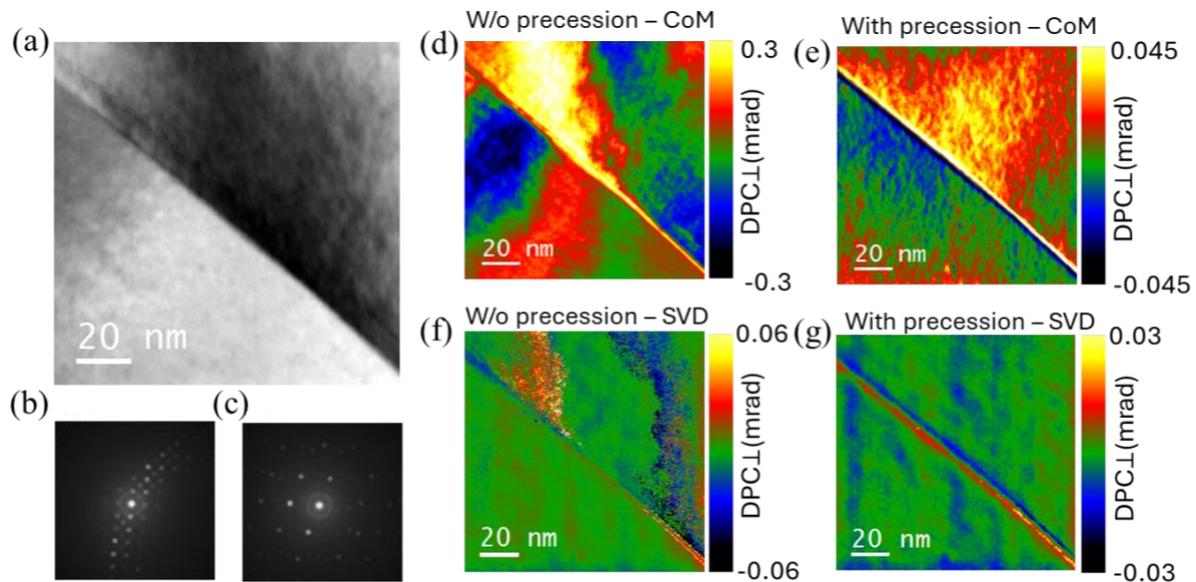

**Figure 3. Quantitative 4D-STEM analysis of a random grain boundary.** (a) ADF-STEM image of the analyzed area containing a grain boundary. (b, c) Representative local diffraction patterns acquired from the two grains on either side of the boundary. (d, e) DPC signal perpendicular to the GB obtained using the CoM method: (d) without beam precession and (e) with beam precession. (f, g) DPC signal perpendicular to GB obtained by the SVD-based disk fitting method: (f) without beam precession and (g) with beam precession.

Figure 3 shows the DPC analysis across a random grain boundary using 4D-STEM. Figure 3a presents the ADF-STEM image of the analyzed region, in which the grain boundary is clearly resolved as a diagonal interface separating two adjacent grains. The local diffraction patterns taken from the two sides of the grains are shown in Figures 3b and 3c. These patterns confirm a random crystallographic misorientation between the neighboring grains.

Figures 3d and 3e show the DPC signal component perpendicular to the grain boundary, reconstructed using the CoM method without and with beam precession. The CoM result without precession exhibits strong spatial variations over both grains, indicating that the measured DPC signal is strongly affected by diffraction-induced intensity asymmetry within the bright-field disk. With beam precession, the broad orientation-dependent contrast is suppressed. However, noticeable background modulation remains throughout the field of view, showing that precession alone is not sufficient to fully remove diffraction-induced contributions from the CoM-based reconstruction as discussed before for polycrystalline samples.

Figures 3f and 3g show the DPC maps of the signal component perpendicular to the grain boundary obtained using the SVD-based disk fitting method without and with beam precession. Compared with the CoM-based results, the SVD-based maps are substantially more uniform within the grains and show much clearer localization of the DPC contrast at the grain boundary and better uniformity along the grain boundary. The improvement is most evident for the combination of precessiond and SVD-based disk fitting in Figure 3g, where the grain boundary is resolved as a narrow high-contrast feature on a fairly homogeneous background.

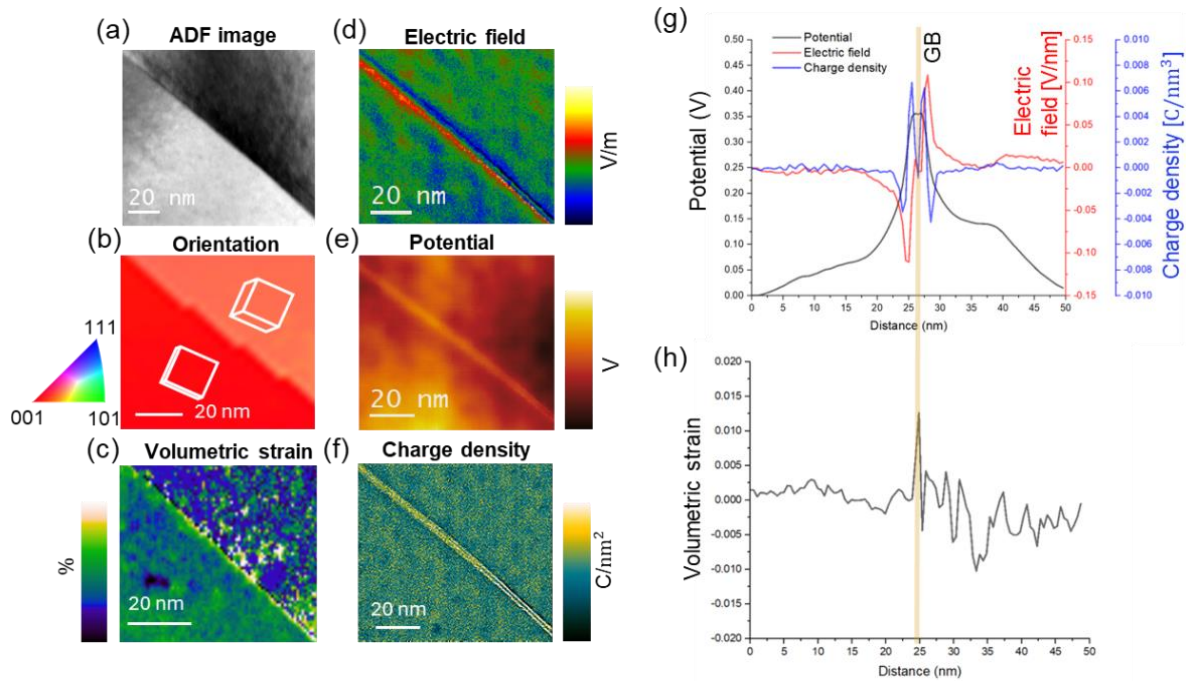

**Figure 4. Quantitative electrostatic potential, electric-field and charge-density analysis across a single random grain boundary using precession 4D-STEM.** (a) ADF-STEM image of the analyzed region. (b) 4D-STEM ACOM orientation map indicating the two neighboring grains; the overlaid cuboids schematically represent their crystallographic orientation. (c) Volumetric strain map of the same area. (d) In-plane electric-field map showing the component normal to the grain boundary obtained by DPC analysis. (e) Electrostatic potential map derived from the measured electric field. (f) Charge-density map calculated from the electric-field distribution. (g) Line profiles of electrostatic potential (black), electric field (red), and charge density (blue) across the grain boundary, with the shaded band marking the grain-boundary position. (h) Volumetric strain profile across the same line.

Figure 4 demonstrates that the present 4D-STEM approach enables correlative analysis of structural and electrostatic quantities across a single grain boundary from the same region. Figure 4a shows the ADF-STEM image of the analyzed region. Figure 4b shows the corresponding 4D-STEM ACOM crystal orientation map, confirming that the interface separates two crystallographically distinct grains; the overlaid cuboids schematically indicate their crystallographic orientation. Figure 4c shows the volumetric strain map of the same area, providing structural information that can be compared directly with the electrostatic response within the same region. Figure 4d shows the in-plane electric-field map of the component normal to the grain boundary, obtained from the precessed-SVD DPC analysis. A pronounced field contrast is observed across the interface, indicating a localized electrostatic response associated with the grain boundary.

Starting from the measured electric-field distribution, additional electrostatic quantities were derived quantitatively. The electrostatic potential shown in Figure 4d was obtained by numerical integration of the electric-field profile along the direction normal to the grain boundary, using $V(x) = -\int E(x)dx$. In practice, the potential was calculated from the corrected electric-field data by cumulative trapezoidal integration and then referenced to the initial point to define the potential offset. The charge-density map in Figure 4f was derived from the spatial gradient of the electric field using Gauss's law, $\rho(x) = \varepsilon_0 \varepsilon_r dE/dx$, followed by conversion into units of $C/nm^3$. In the present calculation, $\varepsilon_r = 1$ was used so that the resulting profile represents the effective charge density directly associated with the measured field gradient.

The line profiles in Figure 4g summarize the correlated electrostatic behavior of the SCL across the grain boundary. The electric-field profile shows a strong divergence on the two sides of the boundary, consistent with the presence of an interfacial electrostatic field associated with a space-charge region. In addition to this broader interfacial response, a narrow feature is observed at the grain-boundary center. Because this central contribution is spatially confined to the boundary core, it should mostly be attributed to the local mean inner potential contribution of the GB rather than the extended space-charge field itself. The corresponding electrostatic potential reconstructed from the measured field shows a positive potential barrier at the GB rather than the expect reduced mean inner potential due to the lower atomic density at the GB. This difference can partially be explained by the SVD-based fitting accuracy being lowest at the GB and because the beam diameter convolutes the measurement around the GB with the SCL features directly next to it. The potential profile for the SCL outside the GB is not perfectly symmetric about the boundary. This asymmetry is presumably due local charging induced by the electron beam during acquisition, which can distort the absolute balance of the reconstructed potential while preserving the overall interfacial trend.

The charge-density profile shows positive peaks at the boundary position, while negative charge accumulation is observed adjacent to the core, consistent with charge compensation in the surrounding space-charge region. The sharp downward peak at the grain-boundary core arises from the local change in atomic density at the core. In contrast, the volumetric strain profile in Figure 4h does not show a clear spatial correlation with the electrostatic features, indicating that the dominant DPC contrast in this region is not primarily driven by the strain gradient. These results show that, together with the atomic structure and crystallographic information obtained from the same 4D-STEM dataset, the measured electric-field profile can be analyzed to obtain the corresponding electrostatic potential and charge-density distributions across the grain boundary.

However, while systematic errors due to crystal orientation variations and thickness are strongly reduced with the combination of precession and SVD-based fitting, the absolute magnitude of the reconstructed electric field, potential, and charge density are influenced by mean inner potential variations at the grain boundary, segregation-induced variations in mean inner potential within the SCL, surface charging of the cross-sectional TEM specimen, and effects associated with the damaged surface layer introduced during TEM sample preparation. These factors need to be considered when making a fully quantitative interpretation of the reconstructed electrostatic profiles. The effects of surface charge and the damage layer are difficult to estimate quantitatively and vary depending on preparation and imaging conditions, therefore we focus here on the influence of the mean inner potential at the GB and within the SCL.

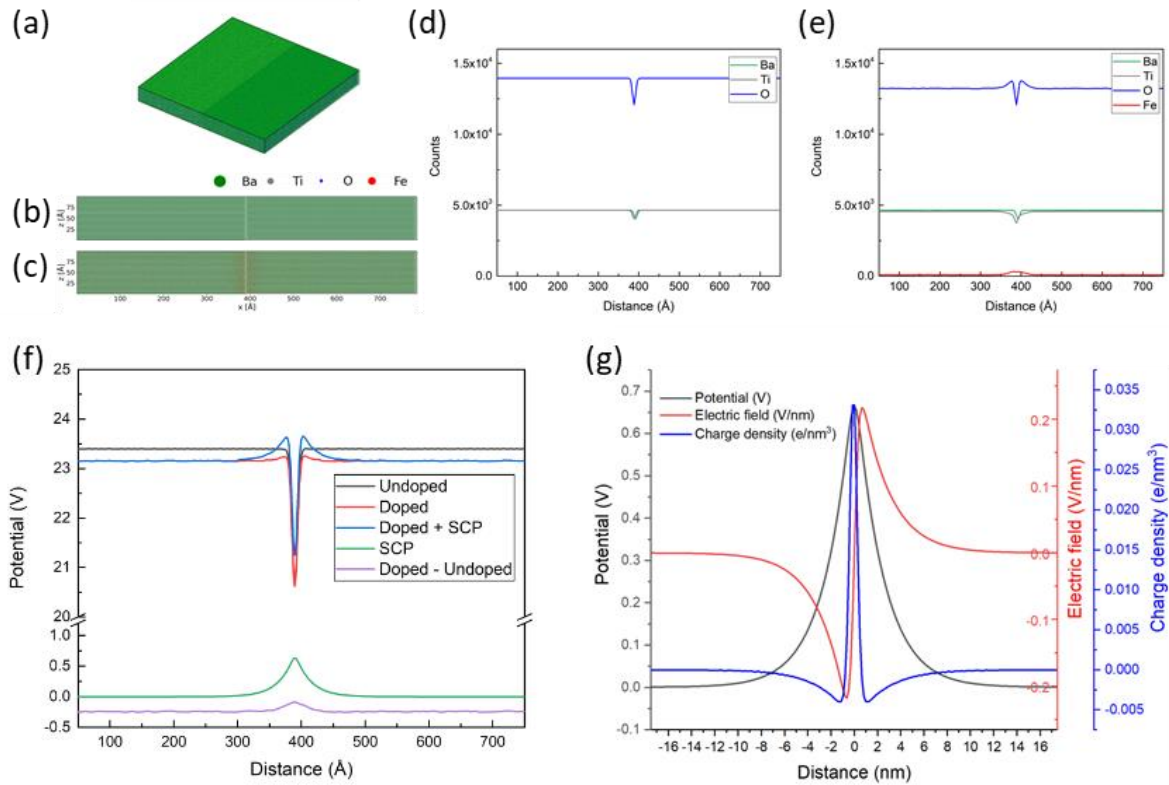

**Figure 5. Atomic structure and electrostatic analysis of a Σ5 grain boundary in BaTiO₃ based on atomistic modeling.** (a) Atomistic simulation cell of the grain-boundary model. (b, c) Atomic configurations of the grain-boundary region for the undoped and Fe-doped models. (d, e) Element-resolved atom counting profiles across the grain boundary for the undoped and the Fe-doped models. (f) Comparison of the mean inner potential (MIP) profiles for the undoped and doped models together with the space-charge potential (SCP), the combined potential, and the difference between the doped and undoped MIP profiles. (g) Electrostatic potential, electric field, and charge-density profiles reconstructed for the pure SCL model across the grain boundary.

Figure 5 provides an atomistic-model-based framework for interpreting the DPC signal across a Σ5 grain boundary in BaTiO₃ as exemplary case study using the experimentally measured compositional profiles for the Fe distribution. Figures 5a-c show the grain-boundary model and the corresponding undoped and Fe-doped atomic structures, while Figures 5d and 5e show the associated elemental profiles. Based on these profiles, the MIP contribution and the space-charge potential were evaluated separately and then combined, as summarized in Figure 5f. The results show that the gradient of the segregation-induced MIP changes contribute to the measured DPC signal and therefore must be considered for a quantitative interpretation of the electric field in the SCL. In the present case, the space-charge contribution is noticeably larger than the MIP contribution within the SCL, with a potential variation of about 0.7 V compared with about 0.2 V from the MIP term, indicating that the electric field of the space-charge-layer signal is the dominant contribution in the SCL. Directly at the GB core, segregation and atomic density induced MIP changes dominated over the electrostatic potential of the SCL as discussed previously [30] preventing an accurate interpretation in terms of the electrostatic profile of the SCL. Figure 5g shows the electrostatic potential, electric-field, and charge-density profiles reconstructed for the combined model in good agreement with the experimental data. These profiles indicate that the experimentally measured DPC signal should be understood as the sum of the electric field response of the SCL and the chemically induced MIP contribution near the grain-boundary core. Depending on the segregating species, the MIP gradient may either enhance or reduce the apparent DPC amplitude. Thus, reliable quantitative analysis of grain boundary DPC data requires explicit consideration of both space-charge electrostatics and segregation-induced MIP variations. The present model does not represent the exact experimental grain boundary type and does not account for possible surface-damage effects, nevertheless, the simulated electrostatic potential and electric-field distributions show reasonably good semi-quantitative agreement with the experimentally measured values. This agreement supports the interpretation that the main features of the experimentally measured DPC response are due to the space-charge layer electrostatics.

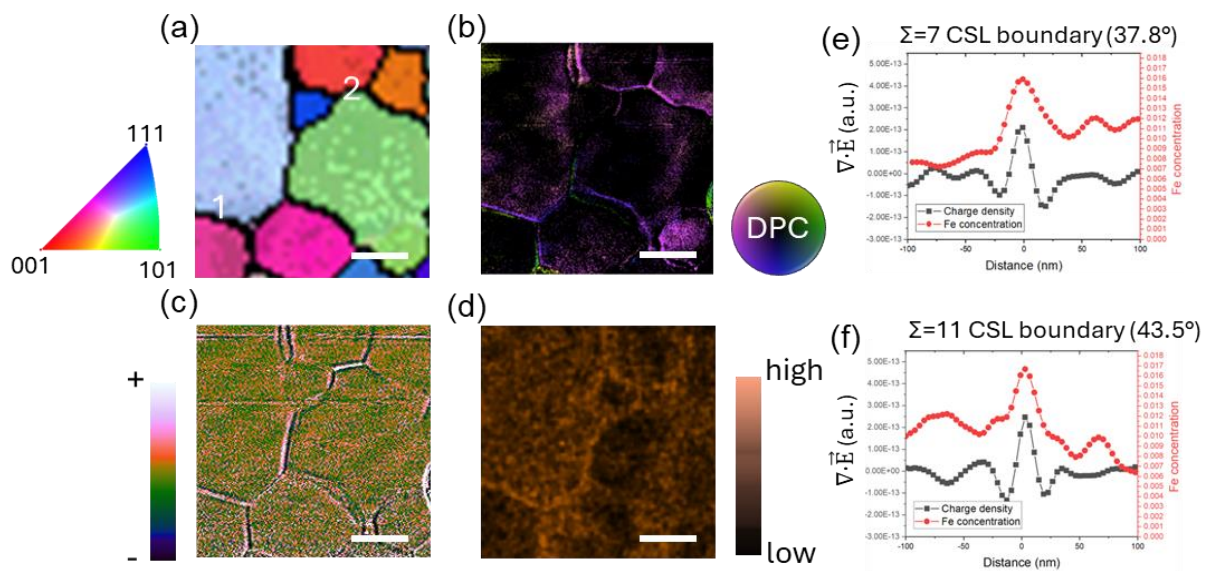

**Figure 6. Large-area correlative mapping of microstructure, DPC response, reconstructed charge density, and Fe distribution in polycrystalline Fe-doped SrTiO3.** (a) ACOM orientation map of the analyzed region, with the orientation color key shown on the left; the two grain boundaries selected for detailed analysis are labeled 1 and 2. (b) Corresponding large-area DPC map, with the DPC color wheel shown as a reference, revealing pronounced interfacial contrast at the grain boundaries. (c) Charge-density map reconstructed from the divergence of the DPC-derived electric field, showing localized electrostatic features at the grain boundaries while the grain interiors remain comparatively uniform. (d) Fe-concentration map of the same region, highlighting spatial variations in dopant segregation. (e) Line profile across grain boundary 1, comparing the reconstructed charge-density signal with the local Fe concentration. (f) Line profile across grain boundary 2, again comparing the reconstructed charge-density signal with the local Fe concentration. In both cases, the grain boundaries show distinct interfacial responses, and the charge-density maxima broadly coincide with Fe-enriched regions, although the detailed profile shape and asymmetry vary from one boundary to another. Together, these

results demonstrate the utility of the present approach for large-area correlative analysis of electrostatic and chemical inhomogeneity in polycrystalline oxides. All scale bars correspond to 200 nm.

The main strength of our approach lies in the reduction of the crystallographic (phase, orientation, strain) contributions to the measured DPC signal, which opens the possibility for reliable large area mapping of polycrystalline materials and to perform a correlative analysis directly connecting ACOM based microstructure analysis (local crystallographic orientation, grain boundary type) with electric-field and charge-density mapping within the same large area. This is illustrated in Figure 6 for polycrystalline Fe-doped $SrTiO_3$, which combines an ACOM orientation map with the corresponding DPC, reconstructed charge-density, and Fe-concentration maps. Both the DPC and charge-density maps show only weak variations within the grain interiors, whereas distinct interfacial signals are clearly resolved at the grain boundaries. The Fe map further reveals that dopant segregation is spatially heterogeneous and tends to be enhanced at selected interfaces. Line profiles extracted across two representative grain boundaries imaged mostly edge on show that the reconstructed charge-density maxima broadly coincide with Fe-enriched regions at the GB, supporting a close correlation between local chemistry and electrostatic response. At the same time, the detailed profile shapes are not identical and, in some cases, exhibit pronounced asymmetry, suggesting that the local electrostatic behavior is influenced not only by segregation, but also by the specific grain-boundary structure with contributions from projection effects and possible charging/surface-damage contributions. Such asymmetric profiles have been shown to depend on grain boundary migration during grain growth by phase-field simulations and are often observed at grain boundaries separating large and small grains [38, 42]. While a full statistical correlation and analysis is still challenging, our new approach now gives access to reliably measure the electrostatic properties and combine this with structural and chemical properties in polycrystalline ceramics towards a more comprehensive understanding of complex non-conducting and semiconducting polycrystalline materials complementing indirect bulk measurements such as electrical impendence spectroscopy [40].

## 4. Summary

We have demonstrated that accurate large area electric-field mapping of the space-charge layer around grain boundaries in polycrystalline oxides can be significantly improved by combining electron beam precession with SVD-based bright-field disk fitting in 4D-STEM. This combined approach significantly suppresses diffraction- and orientation-induced artifacts compared to conventional CoM-based analysis and enables more reliable extraction of DPC signals in structurally heterogeneous regions. Using this workflow, we obtained grain boundary resolved electric-field maps and further derived electrostatic potential and charge-density distributions from the measured field profiles for $BaTiO_3$ and $SrTiO_3$. Atomistic modeling of a Σ5 $BaTiO_3$ grain boundary confirmed that the electric field of the SCL is the dominant contribution to the DPC measurement, but showed that segregation-induced mean inner potential gradients modify the apparent DPC response and must be considered for a quantitative interpretation. Overall, this study provides an effective strategy for reliable electrostatic mapping in complex polycrystalline materials and extends the applicability of 4D-STEM DPC to the analysis of space charge layer related grain boundary phenomena.

## Data availability

The data generated in this study are available upon request from the authors.

## Acknowledgments

S.K. and C.K greatly appreciate the discussions with Roger A. De Souza on space charge layers in oxide ceramics.

## Funding Information

The authors express their gratitude to the Karlsruhe Nano Micro Facility (KNMFi) for providing access to FIB and TEM facilities. S.K. and C.K. acknowledge the support from collaborative research center FLAIR (Fermi level engineering applied to oxide electroceramics), which is funded by the German Research Foundation (DFG), project-ID 463184206 – SFB 1548. X. M. acknowledges financial support from National Natural Science Foundation of China (127000-832011). This work was supported by the Technology Innovation Program (RS-2024-00418991) funded By the Ministry of Trade, Industry & Energy (MOTIE, Korea). Additionally, the authors appreciate the support received from the Joint Lab Model Driven Materials Characterization (MDMC) and acknowledge backing from the Helmholtz Imaging Project (HIP) BRLEMM.

## Author contributions

C.K. initiated the project. S.K. developed the methods and performed the TEM experiments. S.K., H.C., X.M., M.T., J.K.L., M.S., C.K., and K.A. conducted the simulations and analyzed the simulated data. S.K., H.B., D.W., X.M., A.R.N., Z.D., W.R., R.D.S., K.W., B.-X.X., A.K., and C.K. analyzed and discussed the data. A.K. also contributed to method development, sample provision, and scientific discussion. S.K., X.M., and C.K. wrote the draft. All authors contributed to the revision of the manuscript.

## Competing interests

The Authors declare no competing interests.

**Supplementary Information**

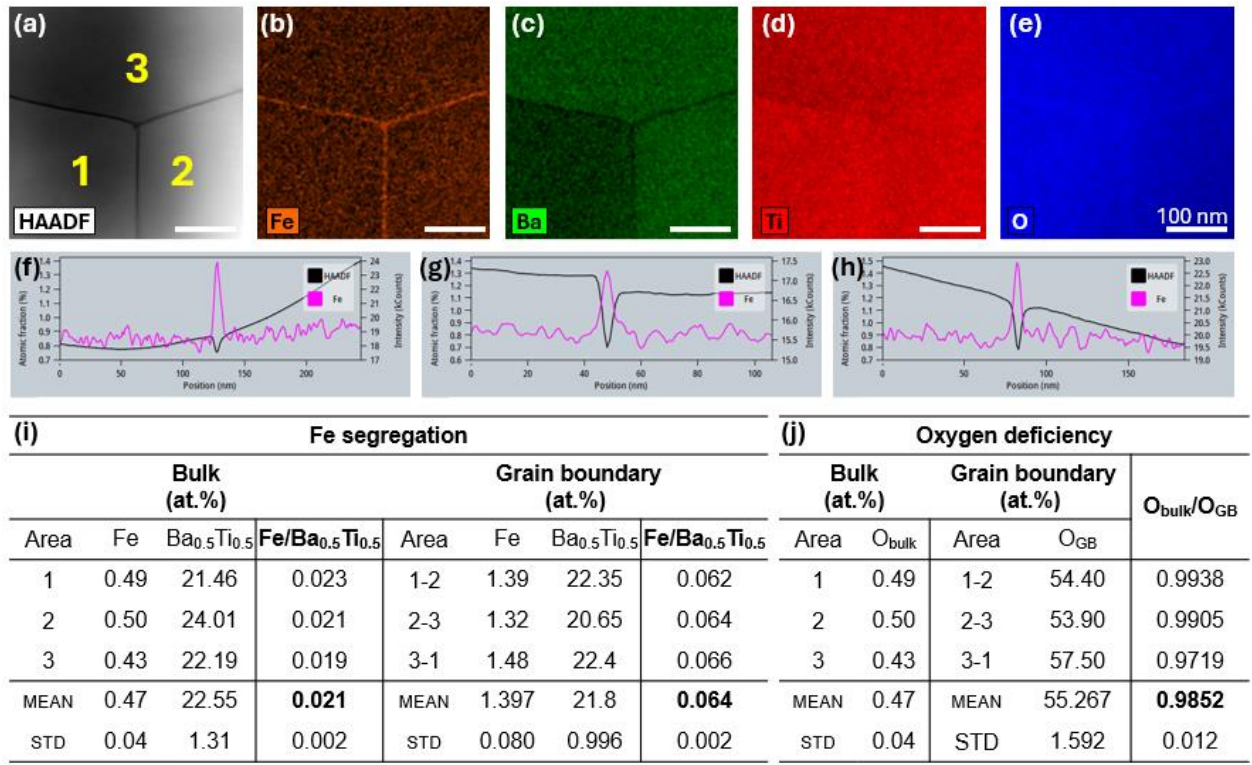

**Figure S1. Experimental STEM-EDS compositional analysis used as reference to construct the Fe-doped grain-boundary model.** (a) HAADF-STEM image of the analyzed region, with three representative grains marked for comparison. (b-e) Corresponding STEM-EDS elemental maps of Fe, Ba, Ti, and O showing Fe enrichment at the grain boundary relative to the adjacent grain interiors. (f-h) Line profiles extracted across the three boundaries, showing the spatial correlation between HAADF intensity and the Fe atomic fraction, highlighting localized Fe segregation with a width of approximately 10 nm around the grain boundary. (i, j) Quantified average compositions and Fe/Ti ratios for the grain-boundary and bulk regions. These experimentally measured compositional profiles were used to define the Gaussian spatial weighting function for Fe substitution on Ti sites in the atomistic model, reproducing the measured Fe/Ti ratios of 0.064 at the grain boundary and 0.021 in the bulk. The experimentally observed oxygen distribution was further used to introduce oxygen deficiency through probabilistic O removal, yielding a bulk-to-grain-boundary oxygen ratio of 0.9852. The resulting composition-informed model served as the basis for the subsequent atomistic simulations.

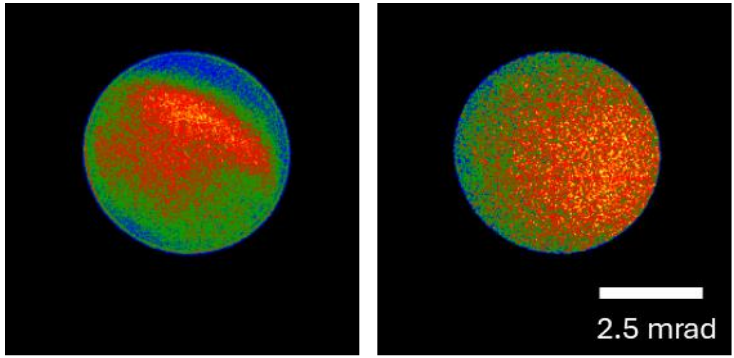

**Figure S2. Effect of beam precession on the bright-field diffraction disk.** Representative diffraction disks acquired with precession off (left) and precession on (right). With beam precession, the strong diffraction contrast is substantially suppressed, and the disk intensity becomes more azimuthally uniform, thereby enhancing the sensitivity to weak electrostatic beam deflections in precession-assisted 4D-STEM.

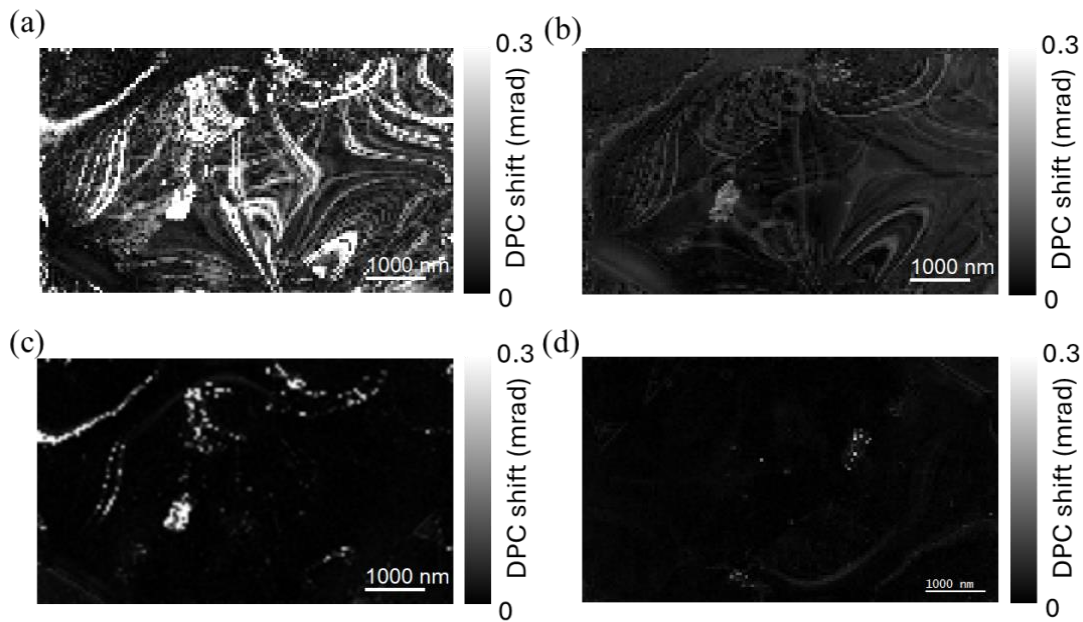

**Figure S3. Amplitude maps of the DPC signal under different 4D-STEM acquisition and processing conditions.** (a, b) DPC amplitude maps obtained using the conventional CoM-based analysis without and with beam precession. (c, d) Corresponding DPC amplitude maps obtained using the enhanced SVD-based disk-center fitting without and with beam precession. In all panels, only the DPC amplitude is shown, highlighting the reduction of diffraction-related artifacts by combining precession and SVD-based disk-center fitting. Scale bars, 1000 nm.

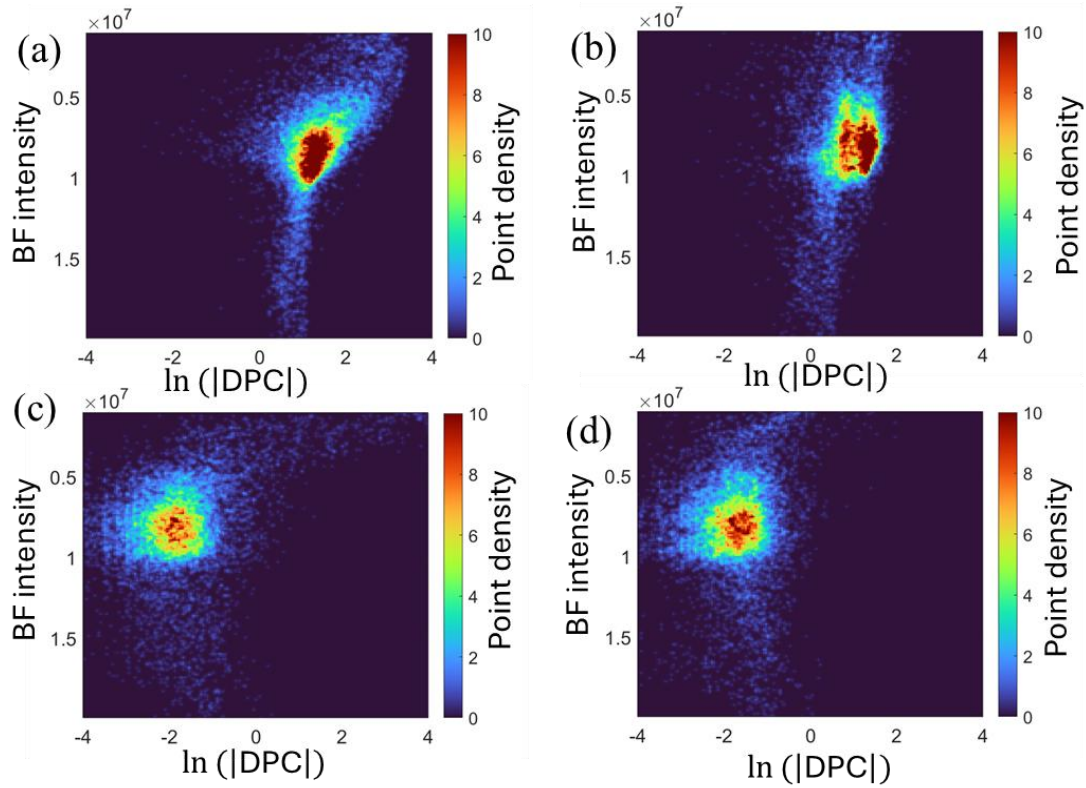

**Figure S4. Two-dimensional histograms of bright-field intensity versus DPC amplitude extracted from the datasets shown in Figure 2.** (a, b) Binned scatter plots of BF intensity versus ln(|DPC|) obtained using the conventional CoM-based analysis without and with beam precession, respectively. (c, d) Corresponding histograms obtained using the enhanced SVD-based disk-center fitting without and with beam precession, respectively. The color scale represents point density. Compared with the CoM-based results, the enhanced SVD-based analysis shows a reduced correlation between BF intensity and DPC amplitude, indicating improved suppression of diffraction-induced contrast artifacts